# Ferromagnetism and Wigner crystallization in Kagome graphene and related structures

Yuanping Chen,[1] Shenglong Xu[2], Yuee Xie,[1] Chengyong Zhong,[1]

Congjun Wu[3] and S. B. Zhang[4]

[1]Department of Physics, Xiangtan University, Xiangtan 411105, People's Republic of China
[2] Condensed Matter Theory Center and Joint Quantum Institute, Department of Physics, University of Maryland, College Park, Maryland 20742, USA
[3] Department of Physics, University of California, San Diego, California 92093, USA
[4]Department of Physics, Applied Physics, and Astronomy Rensselaer Polytechnic Institute, Troy, New York, 12180, USA

Interaction in a flat band is magnified due to the divergence in the density of states, which gives rise to a variety of many-body phenomena such as ferromagnetism and Wigner crystallization. Until now, however, most studies of the flat band physics are based on model systems, making their experimental realization a distant future. Here, we propose a class of systems made of real atoms, namely, carbon atoms with realistic physical interactions (dubbed here as Kagome graphene/graphyne). Density functional theory calculations reveal that these Kagome lattices offer a controllable way to realize robust flat bands sufficiently close to the Fermi level. Upon hole doping, they split into spin-polarized bands at different energies to result in a flat-band ferromagnetism. At a half filling, this splitting reaches its highest level of 768 meV. At smaller fillings, e.g., when $\nu = \frac{1}{6}$, on the other hand, a Wigner crystal spontaneously forms, where the electrons form closed loops localized on the grid points of a regular triangular lattice. It breaks the translational symmetry of the original Kagome lattice. We further show that the Kagome lattices exhibit good mechanical stabilities, based on which a possible route for experimental realization of the Kagome graphene is also proposed.

# Introduction

Carbon is one of the most abundant elements in the universe. To date, a large number of carbon allotropes have been found or synthesized, including fullerene, carbon nanotube, and graphene [1-3]. Graphene is by far the most celebrated carbon material [4], for its excellent properties such as an extremely-high carrier mobility [5,6] and thermal conductivity [7], and mechanical strength [8] with enormous potentials for applications [9-15]. On the other hand, the Coulomb interaction energy in graphene is small when compared with the kinetic energy. In a conventional view, this would imply that carbon is not a good candidate for studying the physics of strongly correlated systems. Furthermore, the strength of the spin-orbit coupling in carbon is exceptionally small [16], which constrains its usefulness in studying the topological physics.

On the other hand, flat band systems are important for both strong interactions and topological physics. Here, regardless its absolute value, interaction energy dominates over kinetic energy due to the quench of the latter. The interplay between a band flatness and interaction energy has given rise to a variety of strong correlation phenomena such as a flat-band ferromagetism [17,18] and Wigner crystallization [19] in the context of a Kagome lattice [20] and the $p_x$-$p_y$-orbital bands of honeycomb [21-23] and Lieb lattices [24-26]. In these systems, due to a destructive interference among the hopping processes, the single-particle eigenstates become localized degenerate states at different plaquettes. The Bloch-wave band states are a superposition of these localized states, and hence, are dispersionless. The salient feature of the flat bands has prompted the suggestions to use it to boost superconducting transition temperature [27] and to host topological states such as the Fractional Quantum Hall States. A combination of non-trivial topology and strong correlation may lead to exotic phenomena as a result of fractional excitations of anyons and fractional statistics.

Recently, fractional Chern insulators based on the flat band structure have also been extensively studied in lattice systems in the absence of magnetic Landau levels [28-30].

Kagome lattice is a lattice which can generate flat band. To date, however, most studies on its flat band physics are based on toy models [31-33]. Experimentally, available materials with a Kagome lattice are mostly frustrated magnets [34-39], which are half-filled Mott insulators. Unfortunately, Fermi level ($E_F$) in such materials is distance away from the flat bands. Given the rich bonding chemistry but simple band structure of carbon, e.g., for graphene only $p_z$ states exist near $E_F$, and exceptional stabilities of its various allotropes, one may ask if it is possible to synthesize carbon-based Kagome or Kagome-like lattices and what would be the physics of strong correlation in them.

In this paper, we identify, by first-principles total-energy calculations, a family of two-dimensional (2D) carbon Kagome-like structures (to be termed Kagome graphene/graphyne), which can host flat bands and relevant physical phenomena. The most basic structure is the Kagome graphene made of carbon triangles on a regular honeycomb lattice. Noticeably, flat bands appear near $E_F$ in all these structures. Upon hole doping, the spin degeneracy of the flat bands is spontaneously lifted to result in a flat-band ferromagnetism with a large spin splitting. At a 1/6 filling factor of the flat bands, a Wigner crystallization is found, while at other filling factors, various charge density wave patterns are expected. These interesting physical phenomena are interpreted based on a Hubbard model. The possibility of realizing an anomalous quantum Hall effect, as well as the possible routes to synthesize the Kagome graphene, are also discussed.

**Model and Methods**

The atomic structure of the Kagome graphene is shown in Fig. 1(a). It has twice as many atoms as a regular Kagome lattice, with the two subsets of atoms in Fig. 1(a) denoted by different colors. Should the bond length between two adjacent different-colored atoms "shrinks" to zero, one recovers the regular Kagome lattice. Other members in the family can be obtained by inserting acetylenic dimers between neighboring triangles. For example, a Kagome graphyne containing one dimer between triangles is shown in Fig. 1(b). In the following, we focus on Kagome graphene, but all the physics discussed below equally applies to other structures as well.

The lattice constant of the Kagome graphene is $a$ (= b) = 5.19 Å. The bond length within a carbon triangle is 1.42 Å, while between the triangles is 1.35 Å. The former is very close to graphene, while the latter is between graphene and acetylene. Although the Kagome graphene contains carbon triangles, its cohesive energy of $E_{coh}$ = 8.26 eV/C is comparable to α- and β-graphynes, 8.28 and 8.35 eV per carbon atom, respectively [40]. To further confirm its lattice stability, we calculated the phonon spectra [see Fig. S1(a) in supplementary information (SI)]. No soft phonon mode is found throughout the BZ.

We performed first-principles calculations within the density functional theory (DFT) as implemented in the VASP codes [41]. The potential of the core was described by the projector augmented wave method [42]. The exchange-correlation interaction between the valence electrons was described by the generalized gradient approximation (GGA), using the Perdew-Burke-Ernzerhof (PBE) functional [43]. A kinetic energy cutoff of 600 eV was used. The atomic positions were optimized using the conjugate gradient method, and the energy and force convergence criteria were $10^{-6}$ eV and $10^{-3}$ eV/Å, respectively. In the case of Kagome graphene, an 11×11×1 $k$-point mash was used for the Brillouin zone (BZ) integration.

## Results and discussions

### 1. Band Structure

Figure 2(a) shows the band structure of Kagome graphene. A flat band appears right below $E_F$ with a band width ~0.1 eV. There are two bands connecting the flat band at Γ point. They cross each other at K point, forming a Dirac point but at a considerably higher energy. The flat band is fully occupied while the two Dirac bands are empty. The partial density of states (PDOS) in Fig. 2(b) indicate that these bands can be attributed to out-of-plane carbon $p_z$-orbitals similar to graphene.

The appearance of the almost flat band can be understood by employing a simple tight-binding model. In the momentum space, the Hamiltonian reads

$$H_0 = \begin{pmatrix} H_{AA} & H_{AB} \\ H_{BA} & H_{BB} \end{pmatrix}, \quad (1)$$

where $H_{AA}$ and $H_{BB}$ are the intra-triangle hopping matrices in sublattice A (red) and B (green) in Fig. 1(a), respectively, and $H_{AB}$ and $H_{BA}$ are the corresponding inter-triangle hopping matrices. They are 3×3 matrices of the form:

$$H_{AA}^{ij} = H_{BB}^{ij} = t_1(1 - \delta^{ij})$$

$$H_{AB}^{ij} = \left(H_{BA}^{ji}\right)^* = t_2 \delta_{ij} e^{-i\vec{r}_i \cdot \vec{k}} \quad (2)$$

where $t_1$ and $t_2$ are the intra-triangle and inter-triangle hopping amplitudes, respectively, and $\vec{r}_{1,2,3}$ are three vectors from the center of a triangle in the A sublattice to those of its three nearest-neighbor triangles in the B sublattice. The tight-binding model with $t_1 = 6\ eV$ and $t_2 = 3\ eV$ captures the essential features of the DFT results, as can be seen in Fig. 2(a). This includes the flat band at $E = t_1 - t_2$, the two Dirac bands with Dirac points at K and K', as well as the quadratic touching between the flat band and the Dirac bands at $\Gamma$, which is protected by the group symmetries $D_{6h}$ at $\Gamma$ and $C_{3v}$ at K (K'). On the other hand, one would not expect the model to work well at higher energies, so we will confine our discussion to states below the Dirac points.

The presence of flat bands in Fig. 2(a) is the result of a special structural property of the Kagome graphene lattice, which implies that there exist local single-particle states as the eigenstates of the band Hamiltonian. Consider a real-space state $|a\rangle$, whose charge distribution is depicted in the inset of Fig. 2(a). Although this is a localized state within a hexagonal plaquette, it is an eigenstate of the Hamiltonian in Eq. (1). This happens because of a total destructive interference of the various hopping amplitudes, which prevents the state from leaking out. The Bloch wave states in the flat band are thus linear combinations of the various $|a\rangle$'s localized in different plaquettes. These localized single-particle states are key to the understanding of strong correlation phenomena.

The flat band and localized states $|a\rangle$ are naturally related to $p$ orbitals on the honeycomb lattice. Consider 3 linearly-independent single-particle states $|1\rangle$, $|2\rangle$ and $|3\rangle$, located on each vertex of a triangle. These states can be decomposed into two irreducible representations (IRREP) of the lattice symmetry group: the 1D IRREP contains the state $|1\rangle + |2\rangle + |3\rangle$, analogous to the $p_x$ orbital in graphene; the 2D IRREP contains the state $|1\rangle + \omega|2\rangle + \omega^2|3\rangle$ and its time reversal partner ($\omega = e^{i\,2\pi/3}$), analogous to local $p_x \pm ip_y$ orbitals in graphene. The localized states $|a\rangle's$ are the flat band state, lying fully in the 2D IRREP. In other words, the localized states discussed above are directly mapped to the localized states in the system of $p$ orbitals in the honeycomb lattice [21].

## 2. Flat band ferromagnetism

The interplay between the band flatness and Coulomb interaction gives rise to many novel phenomena such as the flat-band ferromagnetism [17,18]. Although the Stoner mechanism for ferromagnetism [44] captures the physics of exchange interaction, i.e., spin-polarized electrons keep distance with one another due to Fermi statistics to reduce repulsion, it does not take into account the correlation effect. Electrons often remain unpolarized even under very strong interactions since they can still avoid one another by developing highly-correlated wavefunctions [45]. Spin polarization is typically not favored because the kinetic energy cost is often larger than the exchange energy gain. This competition imposes serious challenge to obtain itinerant ferromagnetism. In contrast, when electrons fill in the flat band, the kinetic energy penalty of spin polarization does not exist anymore, hence, the exchange interaction stabilizes the polarized state. To obtain some insights from DFT calculations, we dope the system by reducing the density of valence electrons (hole doping) to make the flat band partially filled, say, $\nu = \frac{1}{2}$, while maintaining the charge neutrality with a compensating homogeneous background charge. In experiments, doping can be achieved by an electrostatic gating. Figure 3(a) shows the band structure of the doped ferromagnetic ground state, which is spontaneous spin polarized. The spin-up and down bands split, with $E_F$ straddles between the upper empty and lower occupied flat bands. The spin splitting is $\triangle E = 768$ meV, which is roughly on the order of the on-site Coulomb interaction. The cohesive energy of the ferromagnetic ground state is 8.24 eV/C, which is about 20 meV/C lower than the non-polarized state. The lattice constant is increased slightly to 5.27 Å, while all the bond lengths are changed to 1.41 Å. Figure S1(b), SI shows the phonon spectrum for the hole-doped Kagome graphene, showing that the structure is stable. It can be further stabilized by applying an external Zeeman field.

## 3. Wigner crystallization due to the flat band

When the filling inside the flat band is less than $\nu = \frac{1}{2}$, Coulomb interaction can further organize the electrons in the flat band, leading to other interesting phenomena. For instance, Wigner crystallization occurs at $\nu = \frac{1}{6}$, which spontaneously breaks the original translational symmetry of the crystal [21]. A new translational symmetry should emerge but entirely driven by the electron degree of freedom, namely, a Wigner crystallization. A simple analysis shows that a $\sqrt{3} \times \sqrt{3}$ supercell, as marked by the dotted brown lines in Fig. 1(a) which is 30°-rotated from the original cell, is consistent with such a filling of the flat band. Owing to the effect of the band folding, the original spin-up flat band splits into three [see Fig, 4(a)], only one of which is fully occupied to result in a charge gap $\Delta_c \sim 90$ meV with the other two completely empty bands. The lattice constant is also increased from 8.99 to 9.53 Å. Figure 4(b) shows the charge density of the occupied flat-band state, revealing that only $\frac{1}{3}$ of the plaquettes are now occupied. Figure 4(c) shows the magnetic moments on atoms. One can see that this Wigner crystal is ferromagnetic with a moment of $0.055\mu_B$ for atoms on the closed loops inside the occupied plaquettes and $-0.021\mu_B$ for atoms connecting the loops.

## 4. Explanation in terms of the Hubbard model

To understand the above results, we employ the following model of interacting electrons to describe the Kagome-graphene structure,

$$H = H_0 + U \sum_i n_{i,\uparrow} n_{i,\downarrow} + V \sum_{\langle i,j \rangle} (n_{i,\uparrow} + n_{i,\downarrow})(n_{j,\uparrow} + n_{j,\downarrow}), \tag{3}$$

where $H_0$ is the non-interacting tight-binding Hamiltonian introduced in Eq. (1); $U$ and $V$

describe the on-site and nearest-neighboring Coulomb repulsions, respectively. As we will demonstrate below, this minimal model captures the essential physics of both flat band ferromagnetism and Wigner crystallization.

We first discuss the ferromagnetism occurring at $\nu = \frac{1}{2}$. At this filling, each of the localized states $|a\rangle$ in different plaquettes is occupied by one electron. If no electron-electron interaction is included, the spin direction of each localized plaquette state is arbitrary, leading to massive degenerate ground states. Because the localized states of neighboring plaquettes overlap on the bonds they share, the direct exchange interaction dominates, which arises from the two-electron exchange integrals of the on-site and nearest-neighboring repulsions, and favors that the two neighboring localized states have the same spin direction. In consequence, the fully polarized spin configuration is selected as the ground state, exhibiting the flat-band ferromagnetism. The key here is that the kinetic energy is quenched in the flat band while the interaction effect is amplified. In Fig. S2(a) of SI, we show the mean-field calculation of the band structure with $U = 2.8$ eV and $V = 1$ eV, which agrees well with the DFT results.

The ferromagnetism persists at lower fillings. To understand this, consider the process of filling electrons in the flat band. At a low filling, electrons can avoid each other by occupying the localized states centered at plaquettes disconnected with each other, and the interaction energy is simply zero if only the onsite and nearest neighboring repulsions are taken into account. Each localized state is free to choose between spin up or down. As the filling increases, more plaquette localized states are occupied. Beyond $\nu = \frac{1}{6}$, no more plaquette localized states are available to fill; the occupied flat-band states have to touch each other, so their interactions start to take an effect. The touching between two neighboring plaquettes of opposite spins costs the energy of $\frac{1}{72}U + \frac{4}{315}V$ than that of same spin [see Fig. 3(b)]. Consequently, domains constituted of connected plaquette localized states are polarized, whose length scale increases with $\nu$. As the filling reaches the percolation threshold $\nu = \frac{1}{4}$ [46], the area of the largest connected domains scales with the system area, hence, the global magnetization $M$ is developed, which peaks at $\nu = \frac{1}{2}$.

Now we consider the Wigner crystallization at $\nu = \frac{1}{6}$. This particular filling corresponds to the closest packing configuration of occupied plaquette localized states without touching each other, as depicted in Fig. 4(b). This is an incompressible phase, as any change of the closest packing configuration would result in an energy increase due to the nearest-neighbor repulsion between plaquettes. It spontaneously breaks the original lattice translational symmetry to result in an enlarged $\sqrt{3} \times \sqrt{3} R30°$ supercell. This symmetry breaking is consistent with the DFT results in Fig. 4(a). As a matter of fact, the standard mean field treatment of Eq. (3) also reproduces quantitatively the DFT band structure, as can be seen in Fig. S2(b) in SI. As discussed earlier, if only the on-site and nearest neighboring interaction are considered as in the minimal model, the Wigner crystal at $\nu = \frac{1}{6}$ should be paramagnetic as each plaquette can pick random spin direction without affecting the total energy. However, the realistic system studied by DFT includes hoppings beyond nearest neighbors. In this case, the plaquette wavefunctions are no longer exactly localized, but slightly extend outside as evidenced by the small dispersion of the band in Fig. 4(a), such that different plaquette wavefunctions overlap to generate exchange interaction in favor of ferromagnetism.

Note that the DFT calculation and the discussion above are limited to $\nu = \frac{1}{6}$. As Coulomb interaction is a long-range interaction, one can expect that at lower than $\frac{1}{6}$ fillings, Wigner crystals of different patterns should also be stabilized.

Next, we would like to briefly mention the possibility of realizing anomalous Quantum Hall effect in Kagome graphene, whose band structure shows quadratic touches of the flat band with Dirac bands. In general, a quadratic-touching point is unstable in the presence of even an infinitesimal interaction [47], which lifts the degeneracy by opening a gap. In our case at $\nu = \frac{1}{2}$, the spin-polarized flat band spontaneously breaks the time-reversal symmetry in the orbital channel to open a gap. In each triangle, therefore, there is a spontaneously developed non-zero orbital angular momentum $L_z$ around the center of the plaquette, as an order parameter. Consequently, the originally flat band develops a dispersion near the Γ-point, and acquires a non-trivial topological Chern number 1 or $-1$. Using Eq. (3) with the same parameters as before, we obtain $L_z = 0.004\hbar$. Figure S3 in SI shows that a gap $\sim 5$ meV opens up between the flat and Dirac bands, which corresponds to a temperature of about 60 K. Hence, we propose that, at a reasonably low temperature, chiral edge modes should be observed in a Kagome graphene ribbon.

## 5. Experimental feasibility of Kagome graphene

To fabricate the kinetically-stable Kagome graphene, one may consider the following (as detailed in Fig. S4): the bottom line is that the elemental building unit of the triangular carbon rings of Kagome graphene already exists in laboratory as various cyclopropane molecules. One may thus tailor the ligand chemistry of the cyclopropanes to realize a self-assembly of the Kagome graphene, similar to the recent success in self-assembling metastable carbon nanowiggles. In terms of the choice of substrate, the self-assembly process may be carried out on single-layer boron nitride sheet. According to our calculation, the mismatch between a $2\times2$ supercell of boron nitride and the primitive cell of Kagome graphene is smaller than 2.1%. The binding energy between the Kagome graphene and boron nitride is comparable to that of graphene on the same substrate. For more details, one can see Figure S4 in SI, which also shows that the flat band and related phenomena are intact.

## 6. Other related carbon structures

We would like to point out that the aforementioned flat-band physics applies not only to the Kagome graphene, but also to a whole family of related structures, for example, the one in Fig. 1(b) where the triangular rings are linked by acetylenic dimers. We will term such a structure a Kagome graphyne. Here, the lattice constant is $a' = b' = 9.65$ Å, the bond length within the triangular ring is 1.42 Å, within the acetylenic linker is 1.25 Å, and between the ring and linker is 1.34 Å. The phonon spectrum in Fig. S1(c) in SI indicates that Kagome graphyne is also kinetically stable, although its cohesive energy of $E_{coh} = 8.19$ eV/C is lower than that of Kagome graphene, due to the less-stable acetylenic bonds. Figure S5(b) shows that the band structure of Kagome graphyne, with an almost dispersionless flat band right above $E_F$, can be even better to that of Kagome graphene. As results, the spontaneous spin polarization and Wigner crystallization after hole doping are also obtained. When the triangular rings are connected by more acetylenic dimers, a series of carbon structures similar to Kagome graphene such as a Kagome graphdiyne will be obtained. Besides pure carbon networks, covalent organic frameworks such as COF-1 may also be viewed as having a local Kagome structure.

## Conclusion

We show by first-principles calculations that Kagome graphene and related structures are promising artificial materials to study the physics of strong correlation, while traditional carbon-based materials are not. The key is the unusual Kagome structure of carbon atoms, all of which exhibit a flat band right below $E_F$. A tight-binding model is constructed to explain these findings by localized single-particle states as a result of destructive interference among various electron hopping processes. The interplay between band flatness and electron coupling leads to a number of interesting properties such as ferromagnetism, Wigner crystallization, and anomalous quantum Hall effect. Last but not least, we also propose potential synthesis pathways to fabricate the Kagome graphene and related structures.

Recently, Cao et al. reported that bilayer magic-angle graphene superlattice exhibits ultraflat bands near charge neutrality owing to the strong interlayer coupling[48,49]. These flat bands exhibit correlated insulating phases at half-filling, which are not expected in a non-interacting picture. Upon electrostatic doping away from these correlated insulating states, tunable zero-resistance states are observed. These results are complementary with ours, which jointly open a door to explore exotic many-body quantum phases in 2D carbon materials without external magnetic field.

## Acknowledgements

Work in China was supported by the National Natural Science Foundation of China (Nos. 51376005, 11474243, and 11729402), work at RPI was supported by US DOE under Grant No. DE-SC0002623, and work at UCSD was supported by NSF DMR-1410375.

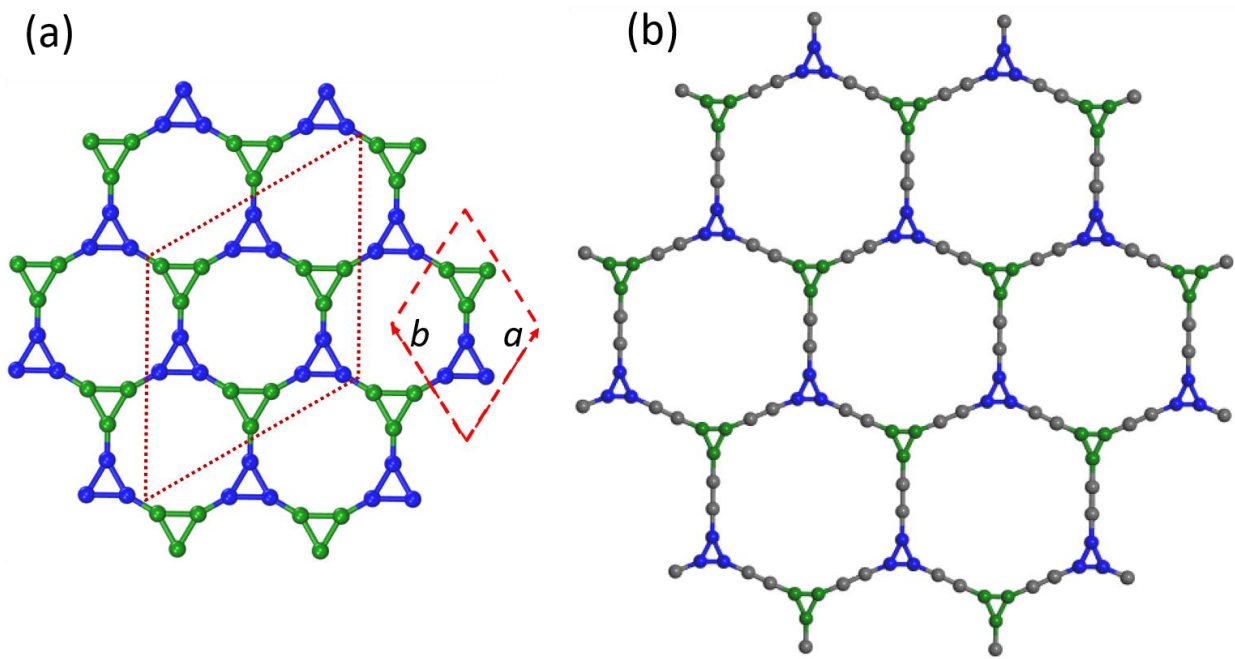

Figure 1 Atomic structure of (a) Kagome graphene, where the primitive unit cell is shown by the dashed red line. Blue and green triangle rings of atoms are two sublattices of the system. Dotted brown line shows a $\sqrt{3} \times \sqrt{3}$ supercell as a result of Wigner crystallization. (b) Kagome graphyne, where the triangular rings are linked by acetylenic carbon linkers.

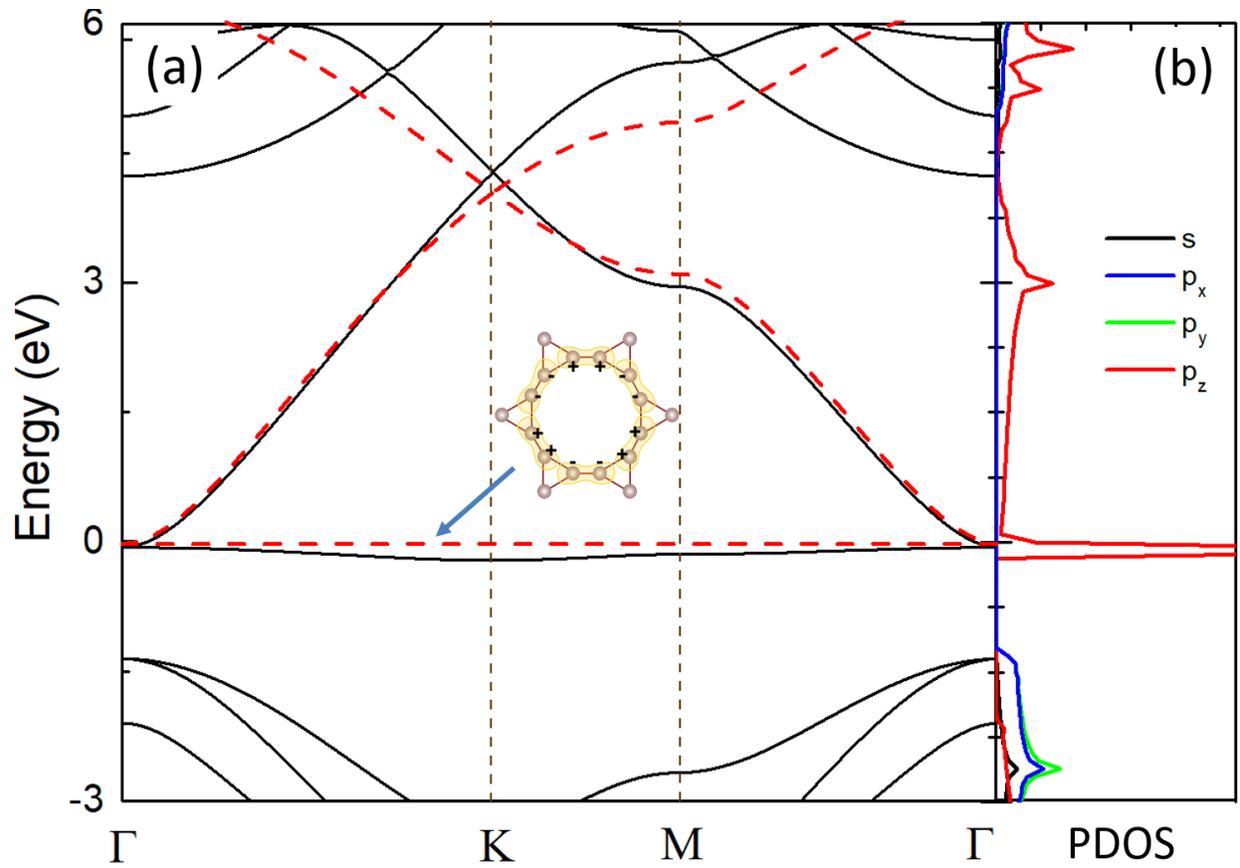

Figure 2 (a) Band structure of Kagome graphene, where the solid and dashed lines are the calculated results of DFT and tight-binding model in Eq. (1), respectively. Inset shows the wave-function for a localized state in the flat band. (b) Partial density of states (PDOS) of Kagome graphene.

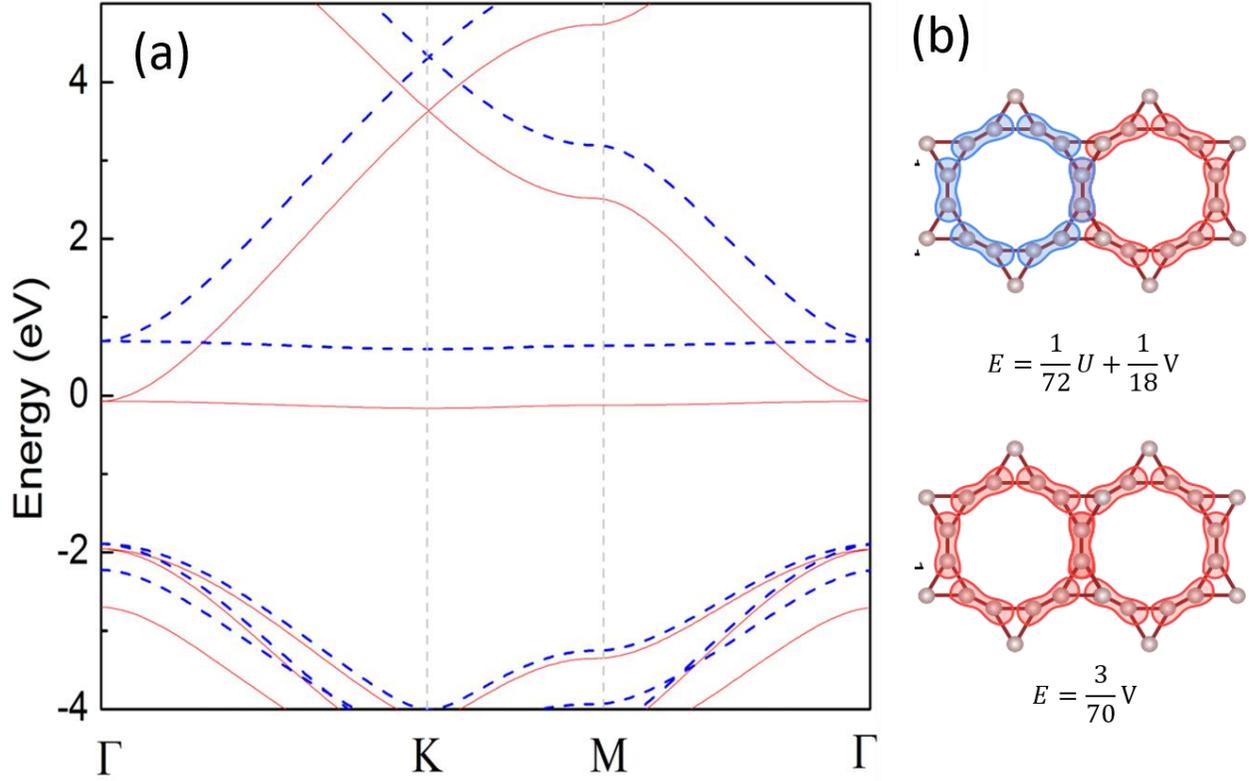

Figure 3. Flat-band ferromagnetism. (a) DFT band structure of Kagome graphene after one-hole doping to result in a ferromagnetic ground state. (b) Wavefunctions of the localized states in the simplified model [cf. Eq. (1)] with calculated energies. The touching of two localized states with opposite spins would cost more energy than that with same spin. This energy cost is the origin for the global ferromagnetism seen here.

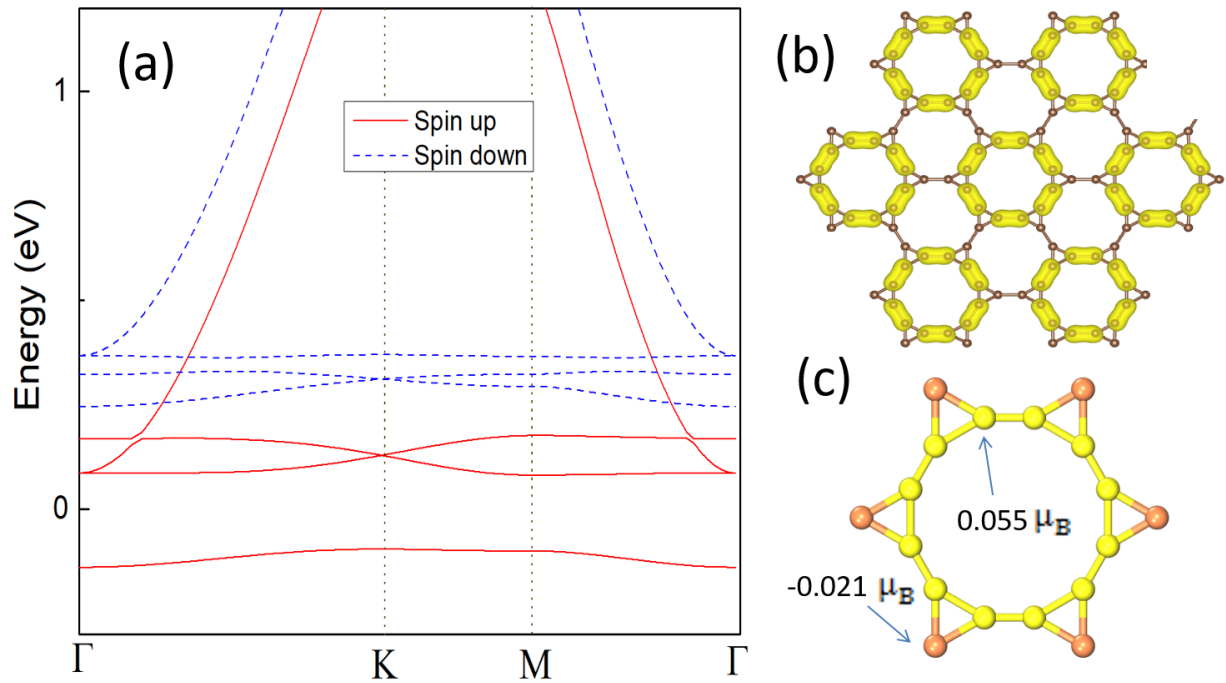

Figure 4 (a) Band structure of Kagome graphene in the $\sqrt{3} \times \sqrt{3}$ supercell doped by 5 holes, i.e., the filling factor is 1/6 for the flat band. (b) Charge density contour for states in the occupied flat band in panel (a). It leads to a Wigner crystallization. (c) Magnetic moments of carbon atoms in the supercell.